\documentclass{pasa}%

\title[Disc formation and fragmentation]{A theoretical perspective on the formation and fragmentation of protostellar discs}
\author[A Whitworth et al.]{A Whitworth$^1$\thanks{email: ant@astro.cf.ac.uk} \and O Lomax$^1$ \\
\affil{$^1$School of Physics \& Astronomy, Cardiff University, Cardiff CF24 3AA, Wales, UK}}

\jid{PASA}
\doi{10.1017/pas.\the\year.xxx}
\jyear{\the\year}

\newcommand{\ga}{\stackrel{>}{\sim}}
\newcommand{\la}{\stackrel{<}{\sim}}
\newcommand{\subB}{_{_{\rm B}}}

\newcommand{\subL}{_{_{\rm L}}}
\newcommand{\subO}{_{_{\rm O}}}
\newcommand{\subR}{_{_{\rm R}}}
\newcommand{\subS}{_{_{\rm S}}}

\usepackage[authoryear]{natbib}
\bibpunct{(}{)}{;}{a}{}{,}
\begin{document}

\begin{abstract}
We discuss the factors influencing the formation and gravitational fragmentation of protostellar discs. We start with a review of how observations of prestellar cores can be analysed statistically to yield plausible initial conditions for simulations of their subsequent collapse. Simulations based on these initial conditions show that, despite the low levels of turbulence in prestellar cores, they deliver primary protostars and associated discs which are routinely subject to stochastic impulsive perturbations; consequently misalignment of the spins and orbits of protostars are common. Also, the simulations produce protostars that collectively have a mass function and binary statistics matching those observed in nearby star formation regions, but only if a significant fraction of the turbulent energy in the core is solenoidal, and accretion onto the primary protostar is episodic with a duty cycle $\ga 3000\,{\rm yr}$. Under this circumstance a core typically spawns between 4 and 5 protostars, with high efficiency, and the lower-mass protostars are mainly formed by disc fragmentation. The requirement that a proto-fragment in a disc lose thermal energy on a dynamical timescale dictates that {\bf there is a sweet spot for} disc fragmentation at radii $70\,{\rm AU}\la R\la 100\,{\rm AU}$ and temperatures $10\,{\rm K}\la T\la 20\,{\rm K}$, and this might explain the Brown Dwarf Desert.
\end{abstract}


\maketitle

\section{INTRODUCTION}\label{SEC:INT}

Discs play a critical role in the formation of stars and planets and moons. They act as repositories for the angular momentum that must be lost by the matter condensing into a star or planet or moon \citep[e.g.][]{ZhuZetal2009, LiBaetal2014}, but they also constitute reservoirs of material from which further stars and/or planets and/or moons can form \citep[e.g.][]{WhitStam2006, Stametal2007b, Chabetal2014}. This paper is concerned with the formation of discs around newly-formed (primary) {\it stars}, and the gravitational fragmentation of such discs to produce additional (secondary) {\it stars}. These {\it stars} can have masses equal to those of {\it planets}, but here they are termed {\it stars} on the grounds that they form early in the lifetime of the disc, by gravitational instability (as opposed to core accretion), on a relatively short dynamical timescale, and -- in the first instance -- with an approximately uniform elemental composition (presumably reflecting the composition prevailing in the local interstellar medium). To put the physics of disc formation and fragmentation in context, we start by discussing the prestellar cores out of which stars and their attendant discs are presumed to condense \citep[e.g.][]{diFretal2007, Wardetal2007}; it is core properties that determine the initial and boundary conditions for disc formation.

In Section \ref{SEC:INF} we discuss {\bf procedures for extracting statistical information about the intrinsic structures of cores from observations of an ensemble of cores.} In Section \ref{SEC:SIM} we present the results of simulations that follow the evolution of an ensemble of cores having properties constrained, as closely as possible, by observations of cores in the nearby Ophiuchus star formation region; we discuss the results, both from the viewpoint of the statistics of the stars and multiple systems formed, and from the viewpoint of the highly chaotic environment influencing the subsequent formation and evolution of the associated circumstellar discs. In Section \ref{SEC:STA} we discuss the statistical relationship between core properties and the stars that they spawn, in order to emphasise further the dynamical and disorganised environments in which circumstellar discs are likely to form and evolve. In Section \ref{SEC:BAS} we present the basic theory of the gravitational fragmentation of a disc. In Sections \ref{SEC:GAM} we calculate how fast a proto-fragment in a circumstellar disc must lose thermal energy in order to condense out of the disc, and in Section \ref{SEC:BRO} we convert this requirement into an explanation for the brown dwarf desert. In Section \ref{SEC:EPI} we explore how cool a circumstellar disc has to be for proto-fragments to condense out by gravitational instability, and conclude that discs are only sufficiently cool if radiative feedback from the central primary star is episodic. In Section  \ref{SEC:ORB} we calculate how fast a proto-fragment in a circumstellar disc must lose angular momentum in order to condense out of the disc, and explain why the critical dynamical timescale for condensation of a proto-fragment in a disc is of order the orbital period of the proto-fragment. Our main conclusions are summarised in Section \ref{SEC:CON}.

\section{INFERRING THE STATISTICAL PROPERTIES OF PRESTELLAR CORES FROM OBSERVATION}\label{SEC:INF}%

In well observed, nearby star formation regions like Ophiuchus, one can identify -- rather unambiguously -- the small fraction of material destined to form the next generation of stars. It is concentrated in dense, low-mass, self-gravitating prestellar cores, which appear to be relatively isolated from one another \citep{Mottetal1998, Andretal2007}. Thus, although an individual core may continue to grow by accreting from the surrounding lower-density gas, it is likely to be finished with forming stars before it interacts significantly with another neighbouring prestellar core. Therefore it is interesting to explore the intrinsic properties of prestellar cores; to use them as the basis for detailed numerical simulations of core collapse and fragmentation (see Section \ref{SEC:SIM}); and to analyse how the properties of cores relate, statistically, to the properties of the stars they spawn (see Section \ref{SEC:STA}).

On the basis of far-infrared and submillimetre continuum intensity maps, one can estimate the dust column-density through, and mean dust temperature along, different lines of sight through a prestellar core. Knowing the distance to the core, one then obtains the core mass, and its projected physical dimensions (i.e. projected area and aspect ratio). On the basis of molecular-line observations, one can also evaluate both the mean gas-kinetic temperature and the non-thermal component of the radial velocity dispersion. Inevitably, one does not know the distributions of density, temperature and radial velocity dispersion along the line of sight, and one has no information on the velocity dispersion perpendicular to the line of sight. Therefore, it is impossible to construct a detailed three-dimensional model of a particular core without making a large number of extremely {\it ad hoc} assumptions.

Since the processes forming cores are chaotic, an alternative procedure is to consider a large ensemble of observed cores. With a sufficiently large number of observed cores, and assuming they are randomly oriented, one can straightforwardly deduce the distributions of mass, size, aspect ratio, temperature, and three-dimensional velocity dispersion, plus any significant correlations between these global parameters \citep[e.g.][]{Lomaetal2013}. A synthetic core (or suite of cores), representing this ensemble, can then be generated by picking values from these distributions and correlations. The large-scale density profile of the synthetic core can be approximated with a critical Bonnor-Ebert profile, or a simple Plummer-like form \citep{WhitWard2001} 
\begin{eqnarray}
\rho(r)&=&\rho\subO\,\left\{1+\left(r/r\subO\right)^2\right\}^{-1}\,.
\end{eqnarray}
The only remaining issue is to define the internal substructure of the synthetic core. This is critical, since it determines how the core subsequently fragments.

The procedure we have adopted is to assume that the internal non-thermal velocity field is turbulent, with power spectrum $P_k\propto k^{-n}$. We have adopted $n=4$, but this is not very critical, as long as $3\la n\la 5$, since most of the power is then concentrated at short wave-numbers. In addition, we must specify the ratio of solenoidal to compressive turbulent energy, and this {\it is} critical, since it strongly influences the properties of the stars that the core spawns, as we discuss in Section \ref{SEC:SIM}. We also adjust the phases of the shortest wave-number solenoidal and compressive modes, i.e. those on the scale of the core, so that they are centred on the core. In other words, these modes are obliged to represent -- respectively -- ordered rotation and ordered radial excursions of the core; all other modes have random phases \citep[e.g.][]{Lomaetal2015b}. Finally we introduce fractal density perturbations, correlated with the velocity perturbations. The fractal dimension is $D\simeq 2.4$ and the density exponent (determining the relative amplitude of the fractal density perturbations) is dictated by the Mach Number of the turbulence.

We emphasise that there is not intended to be -- indeed, there should not be -- a one-to-one correspondence between the individual synthetic cores in our numerically generated ensemble, and the individual observed cores informing the distributions and correlations. They are only related in the sense that we hope they have been drawn from similar underlying distributions. Figure 1 is a grey-scale column-density image of a prestellar core generated in this way. The procedure for generating synthetic cores, and its application to the cores observed in Ophiuchus, is described in more detail in \citet{Lomaetal2013, Lomaetal2014}.

\begin{figure}
\begin{center}
\includegraphics[width=\columnwidth,angle=270]{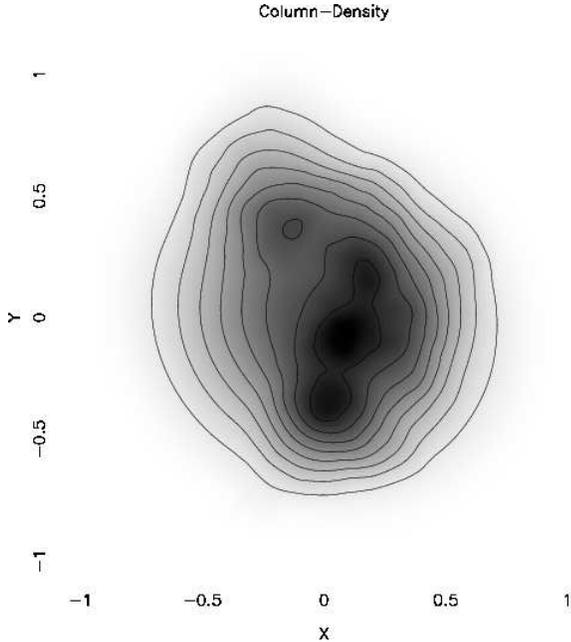}
\caption{Grey-scale column-density image of a synthetic core}\label{FIG:FRACTALCORE}
\end{center}
\end{figure}

\section{SIMULATIONS OF THE COLLAPSE AND FRAGMENTATION OF PRESTELLAR CORES}\label{SEC:SIM}%

We have simulated the collapse and fragmentation of an ensemble of 100 prestellar cores constructed in this way, i.e. 100 prestellar cores that, if placed at the distance of Ophiuchus and observed in dust continuum and molecular line radiation, would be indistinguishable {\it statistically} from the actual cores in Ophiuchus \citep{Lomaetal2015a}. Whether they are a good representation of the cores in Ophiuchus depends on whether the assumptions we have made in the preceding section are valid. The assumptions that turn out to be critical are (i) that the cores are randomly oriented \citep{Lomaetal2013}, (ii) the treatment of radiative feedback \citep[][see below]{Lomaetal2014, Lomaetal2015a}, and (iii) the ratio of solenoidal to compressive turbulence \citep{Lomaetal2015b}. Our assumptions about the density profile, the fractal dimension, and the density- and velocity-scaling exponents appear to be much less critical, in the sense that they do not greatly influence the properties of the stars that form in the simulations. We note that the observed non-thermal kinetic energies in the Ophiuchus cores -- here interpreted as isotropic turbulence -- are typically much smaller than their gravitational potential energies, specifically
\begin{eqnarray}
\alpha_{_{\rm NONTHERM}}&=&\frac{E_{_{\rm NONTHERM}}}{\left|E_{_{\rm GRAV}}\right|}\;\,\stackrel{<}{\sim}\;\,0.1\,.
\end{eqnarray}
The associated non-thermal velocities are typically trans-sonic.

The simulations use the {\sc seren} $\nabla h$-SPH code \citep{Hubbetal2011}, with $\eta =1.2$ (so a particle typically has $\sim 56$ neighbours). Gravitational forces are computed using a tree, and the \citet{MorrMona1997} formulation of time-dependent artificial viscosity is invoked. In all simulations, an SPH particle has mass $m_{_{\rm SPH}}=10^{-5}\,{\rm M}_{_\odot}$, so the opacity limit ($\sim 3\times 10^{-3}\,{\rm M}_{_\odot}$) is resolved with $\sim 300$ particles. Gravitationally bound regions with density higher than $\rho_{_{\rm SINK}}=10^{-9}\,{\rm g}\,{\rm cm}^{-3}$ are replaced with sink particles \citep{Hubbetal2013}. Sink particles have radius $r_{_{\rm SINK}}\simeq 0.2\,{\rm AU}$, corresponding to the smoothing length of an SPH particle with density equal to $\rho_{_{\rm SINK}}$. The equation of state and the energy equation are treated with the algorithm described in \citet{Stametal2007a}, which captures approximately the effects of radiation transport.

Radiative feedback from protostars is also included, using two distinct prescriptions. In the first prescription, labelled {\it continuous feedback}, accretion onto a protostar is presumed to occur at the same rate at which matter is assimilated by the associated sink, so the stellar luminosity is
\begin{eqnarray}
L_\star&\sim&\frac{GM_{_{\rm SINK}}\dot{M}_{_{\rm SINK}}}{3R_{_\odot}}\,.
\end{eqnarray}
Here $M_{_{\rm SINK}}$ is the mass of the sink, $\dot{M}_{_{\rm SINK}}$ is the rate at which the sink is growing due to accretion, and $3R_{_\odot}$ is the approximate radius of a low-mass protostar \citep{PallStah1993}. In the second prescription, labelled {\it episodic feedback}, we use the phenomenological ``sub-sink'' accretion model described in \citet{Stametal2011} -- and also used in \citet{Stametal2012} and \citet{Lomaetal2014}. In this phenomenological model, which is based on the theory developed by \citet{ZhuZetal2009}, highly luminous, short-lived accretion bursts are separated by long intervals of low accretion and low luminosity. During the long intervals, matter collects in the inner accretion disc inside a sink, but the rate of accretion onto the central primary protostar is low -- basically because there is no effective mechanism for redistributing angular momentum: the matter in the inner accretion disc is too hot for gravitational structures to develop and exert torques, and too cool to be thermally ionised and couple to the magnetic field. Consequently the luminosity of the central primary protostar is also low, and the disc cools down and may become cool enough to fragment. Once sufficient matter has collected in the inner accretion disc, it becomes hot enough for thermal ionisation to be significant; the Magneto-Rotational Instability then cuts in, delivering efficient outward angular momentum transport; matter is dumped onto the protostar giving an accretion outburst and an associated peak in the luminosity, which heats the disc and stabilises it against fragmentation.

Each of the 100 cores in the ensemble has been evolved numerically with both continuous and episodic radiative feedback, and with different proportions of solenoidal and compressive turbulence \citep{Lomaetal2014, Lomaetal2015a, Lomaetal2015b}. The simulations reproduce very well the observed mass function of young stars, and their multiplicity properties, including the incidence of high-order hierarchical multiples, but only if at least $\sim 30\%$ of the turbulent kinetic energy is in solenoidal modes, and only if the radiative feedback from protostars is episodic.

Some solenoidal turbulent energy is required, to promote the formation of massive extended circumstellar discs around the first protostars to form, and episodic feedback is required to allow these discs to periodically become sufficiently cool to fragment gravitationally and spawn additional protostars, in particular brown-dwarf protostars.

If there is no solenoidal turbulence, the circumstellar discs around the first protostars to form (which by the end of the simulation are usually the most massive protostars) are compact and unable to fragment. Low-mass protostars still form, by fragmentation of the filaments that feed material towards these more massive protostars, but very few of these low-mass protostars end up below the hydrogen-burning limit, and so in the end there are far too few brown-dwarf protostars to match observations\footnote{The fraction of stars below the hydrogen-burning limit is estimated to be $0.2\;{\rm to}\;0.3$ \citep{Andeetal2011}.}. As the amount of solenoidal turbulence increases, the mean stellar mass decreases, and the fraction of brown dwarfs increases.

Even if there is solenoidal turbulence (and so extended massive circumstellar discs do form around the first protostars), as long as there is continuous feedback, the discs are warm and disc fragmentation is effectively suppressed. Consequently, the first protostars tend to accrete all the mass in their circumstellar discs and become very massive. Again, the upshot is that the mean stellar mass is high and there are very few brown dwarfs. Moreover, the few brown dwarfs that do form are very close to the hydrogen-burning limit, and there are none close to the deuterium-burning limit. Episodic accretion allows circumstellar discs to cool down, and -- provided that the duty-cycle for episodic accretion is sufficiently long, i.e comparable with or greater than the dynamical timescale of the outer parts of the disc, say $\stackrel{>}{\sim}3,000\,{\rm yrs}$ -- the discs are cool enough for long enough to fragment, thereby spawning brown dwarfs and low-mass hydrogen-burning stars. Recent estimates of the duty-cycle for episodic accretion indicate that it may indeed be this long \citep[e.g.][]{Schoetal2013}.

\begin{figure}
\begin{center}
\includegraphics[width=\columnwidth]{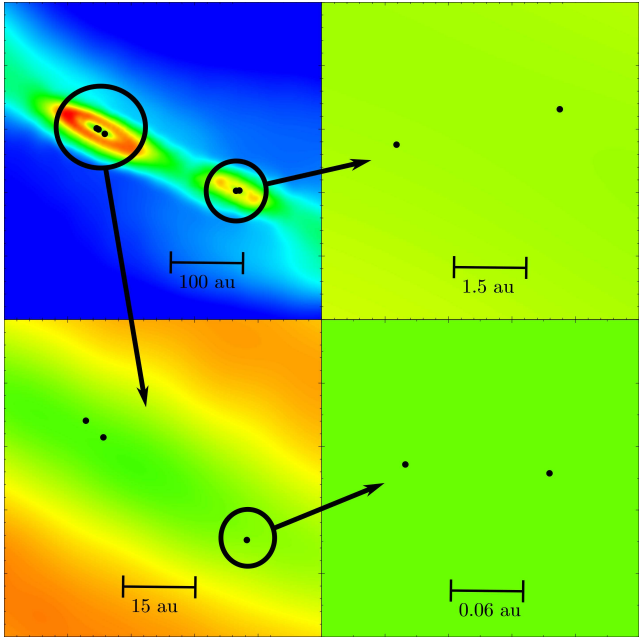}
\caption{Montage of false-colour column-density maps of the central regions of a collapsed core which has formed an hierarchical sextuplet. The protostars are marked with black dots}\label{FIG:SEXTUPLET}
\end{center}
\end{figure}

Disc fragmentation is also conducive to the formation of hierarchical (and therefore stable) higher-order multiples, and our simulations with solenoidal turbulence and episodic feedback deliver distributions of binaries and higher-order multiples that are in good agreement with observation, including systems with up to six components. Figure \ref{FIG:SEXTUPLET} illustrates an hierarchical sextuple system formed in one of our simulations. If there is too little solenoidal turbulence, and/or more continuous radiative feedback, far too few multiple systems are formed.

With continuous radiative feedback, an average core spawns between 1 and 2 protostars; with episodic feedback this increases to between 4 and 5 protostars. This is a controversial result, in the sense that it has usually been assumed that a core typically forms one or two protostars, with the majority of the core mass then being dispersed. However, there are good statistical reasons to suggest that a typical core may well spawn between 4 and 5 protostars \citep[see Section \ref{SEC:STA} and][]{Holmetal2013}. There are also recent observations implying that the number of protostars in a core has been underestimated. Additionally, the fact that young populations include significant numbers of high-order multiples suggests that cores must, at least occasionally, spawn many more than two protostars \citep[e.g.][]{Krauetal2011}. 

A key factor promoting the formation and gravitational fragmentation of discs in simulations with solenoidal turbulence is that the infall onto a circumstellar disc is seldom smooth. The discs form in a very dynamic environment, in which new material is often delivered to a disc quite impulsively, i.e. at an abruptly increased rate, and/or along a filament whose direction and interception point bear little relation to the structure and orientation of the existing disc. Discs are also occasionally perturbed tidally by close passages of other protostars that have formed in the disc, an occurrence that is more frequent if cores spawn between 4 and 5 protostars.

{\bf As well as promoting disc fragmentation, these perturbations represent a stochastic and impulsive input of angular momentum. This has the consequence that discs can be quite poorly aligned with the spins of their central stars, as shown by \citet{Bateetal2010}, and that the spins of stars in multiple systems are often rather poorly aligned with one another and with their mutual orbits \citep[e.g.][]{Lomaetal2015a}.}

\section{THE STATISTICS OF CORE FRAGMENTATION}\label{SEC:STA}%

It has frequently been noted that the prestellar core mass function (hereafter CMF) is similar in shape to the stellar initial mass function (hereafter IMF) -- basically log-normal with a power-law tail at high masses -- but shifted to higher masses by a factor of $4\pm 1$ \citep[e.g.][]{Andretal2010}. This has led to the hypothesis that the shape of the IMF is inherited directly from the CMF. If this inference is correct, then the mapping from the CMF to the IMF must be statistically self-similar. In other words, for a core of any mass $M_{_{\rm CORE}}$, the mean fraction of the core's mass that ends up in protostars ($\eta_{_{\rm O}}$), the mean number of protostars formed from the core (${\cal N}_{_{\rm O}}$), the distribution of {\it relative} stellar masses (which we here characterise as a log-normal with standard deviation $\sigma_{_{\rm O}}$), and the predisposition of these protostars to end up in binaries or higher-order multiples (which we here characterise with an exponent $\alpha_{_{\rm O}}$ such that the relative probability for a protostar of mass $M_\star$ to end up in a binary is proportional to $M_\star^{\alpha_{_{\rm O}}}$) are all universal.\footnote{Thus large $\eta_{_{\rm O}}$ means that a large fraction of the core mass ends up in protostars, large ${\cal N}_{_{\rm O}}$ means that a core typically spawns many protostars, a large $\sigma_{_{\rm O}}$ means that the protostars spawned by a core have a wide range of masses, and a large $\alpha_{_{\rm O}}$ means that there is a strong tendency for the two most massive protostars in a core to be the ones that end up in a binary. The results are not changed significantly if the distribution of relative masses is not lognormal but is characterised by a parameter $\sigma_{_{\rm O}}$ that measures the width of the distribution, or if the relative probability of ending up in a binary depends on mass in a different way but involves a parameter $\alpha_{_{\rm O}}$ such that large $\alpha_{_{\rm O}}$ favours higher-mass protostars, i.e. dynamical biasing \citep{McDoClar1993, McDoClar1995}.} For example, the probability that a core with mass $0.5\,{\rm M}_{_\odot}$ spawns two protostars with masses between $0.1\,{\rm M}_{_\odot}$ and $0.2\,{\rm M}_{_\odot}$ is the same as the probability that a core with mass $5\,{\rm M}_{_\odot}$ spawns two protostars with masses between $1\,{\rm M}_{_\odot}$ and $2\,{\rm M}_{_\odot}$.

\citet{Holmetal2013} have used a Monte Carlo Markov Chain analysis to determine which parameter values deliver an acceptable mapping between the observed distribution of core mass and the observed distributions of stellar mass, binary frequency (as a function of primary mass) and binary mass ratio (as a function of primary mass). This analysis predicts -- very `forcefully' -- that the preferred value of $\eta_{_{\rm O}}$ is $\eta_{_{\rm O}}\simeq 1.0\pm 0.3$, i.e. most of the initial core mass ends up in protostars. We note that $\eta_{_{\rm O}} >1$ is admissible, because, between the time when the mass of a core is measured and the time when it has finished  forming protostars, the core can -- and almost certainly does -- accrete additional matter. The preferred value of ${\cal N}_{_{\rm O}}$ is ${\cal N}_{_{\rm O}}\simeq 4.3\pm 0.4$ (as predicted by the numerical simulations described in Section \ref{SEC:SIM}), so that the shift from the peak of the IMF to the peak of the CMF should be ${\cal N}_{_{\rm O}}/\eta_{_{\rm O}}\simeq 4.3\pm 1.3$ (as observed). The basic reason why ${\cal N}_{_{\rm O}}\sim 4\;{\rm to}\;5$ fits the observations so much better than ${\cal N}_{_{\rm O}}\sim 1\;{\rm to}\;2$ is that the former predicts that the  binary frequency increases monotonically with primary mass, almost exactly as observed (see Figure 3), whereas the latter predicts that the binary frequency decreases with primary mass.

The preferred value of $\sigma_{_{\rm O}}$ is $\sigma_{_{\rm O}}\simeq 0.30\pm 0.03$, so the interquartile range for the protostars spawned by a single core spans a factor of order 2.

Finally, the preferred value of $\alpha_{_{\rm O}}$ is $\alpha_{_{\rm O}}\simeq 0.9\pm 0.6$, meaning that there is some dynamical biasing, i.e. some tendency for the more massive protostars to end up in binaries, but rather weaker than the results of pure $N$-body simulations \citep{McDoClar1995} would suggest. This implies that the protostars in a core do not form first and then pair up. Rather, the identity of a future companion protostar is determined during formation. For example, our simulations of Ophiuchus-like cores suggest that companions to the more massive protostars often form by disc fragmentation.

These aspects of protostar formation, and those discussed in Section \ref{SEC:SIM}, all indicate that the notion of an isolated primary protostar with a symmetric infalling envelope and a symmetric circumstellar disc may be misleading. Circumstellar (and circumbinary) discs probably form in chaotic environments, where impulsive perturbations due to anisotropic lumpy accretion streams and nearby passing protostars are the norm.

\begin{figure*}
\begin{center}
\includegraphics[width=\columnwidth,angle=270]{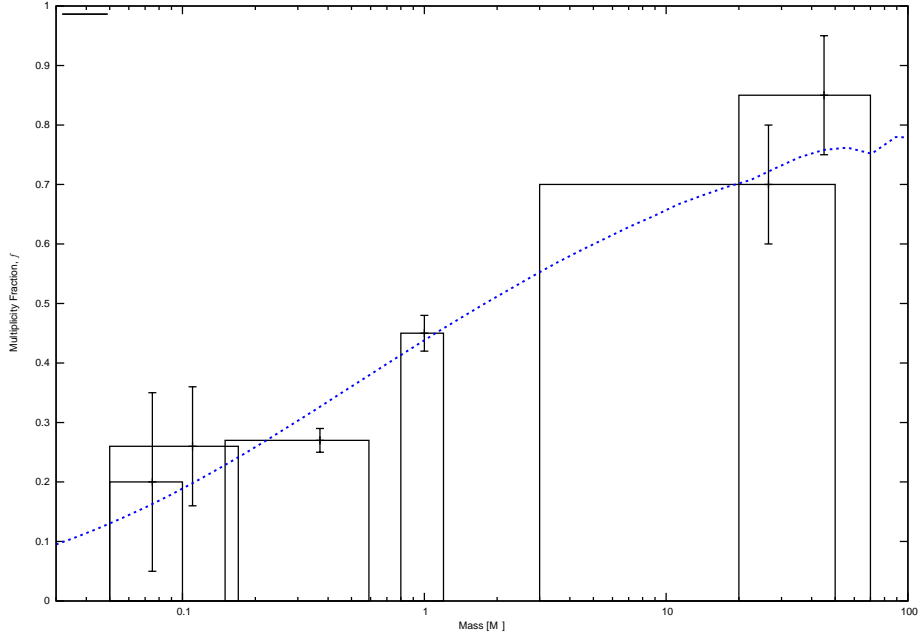}
\caption{The dashed line gives the binary frequency as a function of primary mass {\bf for the best-fitting values of $\eta_{_{\rm O}}\simeq 1.0$, ${\cal N}_{_{\rm O}}\simeq 4.3$, $\sigma_{_{\rm O}}\simeq 0.30$ and $\alpha_{_{\rm O}}\simeq 0.9$ identified by the Monte Carlo Markov Chain analysis}. The boxes represent the observational estimates of multiplicity frequency in different primary-mass intervals, due to \citet{Closetal2003,BasrRein2006,FiscMarc1992,DuquMayo1991,Preietal1999,Masoetal1998}. The error bars represent the observational uncertainties. }\label{FIG:MULTIPLICITYFREQUENCY}
\end{center}
\end{figure*}

\section{THE BASIC THEORY OF DISC FRAGMENTATION}\label{SEC:BAS}%

Consider an equilibrium circumstellar disc with surface density $\Sigma(R)$, isothermal sound-speed $c\subS(R)$, angular speed $\Omega(R)$ and epicyclic frequency $\kappa(R)$. As long as the overall mass of the disc is less than the mass of the central protostar, we can put 
\begin{eqnarray}\label{EQN:OmegaR}
\kappa(R)\;\,\rightarrow\;\,\Omega(R)&\sim&\left(\frac{GM_\star}{R^3}\right)^{1/2}\,,
\end{eqnarray}
(and henceforth the variable $\kappa$ will be used exclusively for opacity). In the sequel we shall not normally include the $R$-dependence of $\Sigma(R)$, $c\subS(R)$ and $\Omega(R)$ explicitly.

Now suppose that a small circular patch of the disc at radius $R$, having radius $r\,\ll\! R$ (i.e. a proto-fragment), becomes unstable and starts to condense out. The radial excursions of this proto-fragment are dictated by the balance between gravitational, pressure and centrifugal accelerations,
\begin{eqnarray}
\ddot{r}&\;\sim\;&-\,2\pi G\Sigma\,+\,\frac{c\subS^2}{r}\,+\,\Omega^2r\,.
\end{eqnarray}
Condensation requires $\ddot{r}<0$, and so the range of unstable proto-fragment radii, $(r_{_{\rm MIN}},r_{_{\rm MAX}})$, is given by
\begin{eqnarray}\label{EQN:rMIN,MAX:1}
r_{_{\rm MIN,MAX}}&\sim&\frac{\pi G\Sigma\,\mp\,\left\{(\pi G\Sigma)^2-(c\subS\Omega)^2\right\}^{1/2}}{\Omega^2}\,.
\end{eqnarray}
There are real roots, and therefore there are only unstable proto-fragments at radius $R$, if $\pi G\Sigma > c\subS\Omega$, i.e.
\begin{eqnarray}\label{EQN:SigmaMIN}
\Sigma&>&\Sigma_{_{\rm MIN}}\;\sim\;\frac{c\subS\Omega}{\pi G}\,.
\end{eqnarray}
This is the Toomre criterion for gravitational instability of an equilibrium disc \citep{Toomre1964}.

We note that proto-fragments with $r<r_{_{\rm MIN}}$ are unable to condense out because their pressure support is stronger than their self-gravity. In contrast, proto-fragments with $r>r_{_{\rm MAX}}$ are unable to condense out because their rotational support is stronger than their self-gravity. 

The timescale on which a proto-fragment condenses out is
\begin{eqnarray}\nonumber
t_{_{\rm COND}}&\sim&\left(\frac{2r}{\ddot r}\right)^{1/2}\\
&\sim&\left\{\frac{\pi G\Sigma}{r}-\frac{c\subS^2}{2r^2}-\frac{\Omega^2}{2}\right\}^{\!-1/2}\,,
\end{eqnarray}
and the fastest condensing fragment has
\begin{eqnarray}\nonumber
r_{_{\rm FAST}}&\sim&\frac{c\subS^2}{\pi G\Sigma}\,, \\
t_{_{\rm FAST}}&\sim&\left\{\frac{(\pi G\Sigma)^2}{2c\subS^2}-\frac{\Omega^2}{2}\right\}^{\!-1/2} \\
&\sim&\frac{t_{_{\rm ORB}}}{2^{1/2}\pi\left\{(\Sigma/\Sigma_{_{\rm MIN}})^2-1\right\}^{\!1/2}}\,,
\end{eqnarray}
where 
\begin{eqnarray}\label{EQN:tORB}
t_{_{\rm ORB}}&=&\frac{2\pi}{\Omega}
\end{eqnarray}
is the orbital period at radius $R$. Thus a proto-fragment can condense out in one orbital period if 
\begin{eqnarray}
\frac{\Sigma}{\Sigma_{_{\rm MIN}}}&\ga&1\,+\,\frac{1}{(2\pi)^2}\;\,\sim\;\,1.025\,;
\end{eqnarray}
this implies that the disc need only be marginally unstable for proto-fragments to condense out on an orbital timescale.

\section{HOW FAST MUST A PROTO-FRAGMENT LOSE THERMAL ENERGY TO CONDENSE OUT?}\label{SEC:GAM}%

In what follows, we shall substitute for the temperature in terms of the isothermal sound-speed, $c_{_{\rm S}}$, viz. 
\begin{eqnarray}
T&\longrightarrow&\frac{\bar{m}\,c\subS^2}{k\subB}\,,
\end{eqnarray}
where $\bar{m}$ is the mean gas-particle mass (which for molecular gas is $\bar{m}\sim 5\times 10^{-24}\,{\rm g}$) and $k\subB$ is Boltzmann's constant. The flux from a blackbody is then
\begin{eqnarray}\label{EQN:FBB}
F_{_{\rm BB}}&=&\sigma\subB\,T^4\;\,=\;\,\frac{2\,\pi^5\,\bar{m}^4\,c\subS^8}{15\,c\subL^2\,h^3}\,,
\end{eqnarray}
where $\sigma\subB$ is the Stefan-Boltzmann constant, and $c\subL$ is the speed of light.

A proto-fragment in a disc can only condense out if it can cool radiatively, on a dynamical timescale. Otherwise it will undergo an adiabatic bounce, and be sheared apart or merge with another proto-fragment. We shall assume that cooling is dominated by thermal emission from dust.

The frequency-averaged mass-opacity for a standard mixture of interstellar gas and dust (at the long wavelengths of concern here) can by approximated by
\begin{eqnarray}\label{EQN:kappaR}
\bar{\kappa}\subR&\sim&\kappa_{_T}\,T^2\;\,\sim\;\,\kappa_{_{c}}\,c\subS^4\,,
\end{eqnarray}
with $\kappa_{_T}\simeq 10^{-3}\,{\rm cm}^2\,{\rm g}^{-1}\,{\rm K}^{-2}$ and $\kappa_{_c}=\kappa_{_T}(\bar{m}/k\subB)^2\simeq 1.2\times 10^{-18}\,{\rm s}^4\,{\rm cm}^{-2}\,{\rm g}^{-1}\,$. The uncertainty on these coefficients is the same as on, say, the mass-opacity that is routinely used to estimate column-densities from $850\,\mu{\rm m}$ intensities, i.e. a factor of order two. For simplicity, we do not distinguish between the Rosseland- and Planck-mean opacities.

At the temperatures with which we are concerned here, $T\la 100\,{\rm K}$, the rotational and vibrational degrees of freedom of a hydrogen molecule are not significantly excited, and therefore the gas is effectively monatomic, with specific internal energy $u=3c_{_{\rm S}}^2/2$. If we assume that the dynamical timescale for a condensing proto-fragment in a disc is the same as its orbital period, then the requirement that a proto-fragment be able to cool on a dynamical timescale becomes
\begin{eqnarray}\label{EQN:GAMMIE1}
\frac{3\Sigma c_{_{\rm S}}^2/2}{t_{_{\rm ORB}}}&\,<\,&\left\{\begin{array}{ll}
2F_{_{\rm BB}}\bar{\tau}\,,\hspace{0.5cm}&\bar{\tau}<1\;{\rm (optically\;thin)}\,,\hspace{1.0cm}\\
2F_{_{\rm BB}}\bar{\tau}^{-1}\,,&\bar{\tau}>1\;{\rm (optically\;thick)}\,,
\end{array}\right.\\\nonumber\\\label{EQN:GAMMIE2}
\bar{\tau}&=&\frac{\Sigma\bar{\kappa}}{2}\,,
\end{eqnarray}
which is equivalent to the Gammie Criterion \citep{Gammie2001}. The Gammie Criterion has been confirmed numerically by various groups, usually using a parametrised cooling law of the form $d\ln [u]/dt=-2\pi/\beta t_{_{\rm ORB}}$, e.g. \citet{Riceetal2003}, although there is still some debate as to {\bf the validity of the Gammie Criterion} \citep[see, for example,][]{MeruBate2011, Riceetal2014}. The Gammie Criterion is basically the same as the Opacity Limit \citep[e.g.][]{Rees1976, LowCLynd1976}, but formulated for the case of two-dimensional fragmentation of a disc (Toomre instability), rather than three-dimensional fragmentation of an extended 3D medium (Jeans instability). We will return to the reason why the dynamical timescale of a proto-fragment might be the same as its orbital timescale in Section \ref{SEC:ORB}.

\section{THE BROWN DWARF DESERT}\label{SEC:BRO}%

\subsection{The optically thin limit}%

If we adopt the first option in Eqn. (\ref{EQN:GAMMIE1}), i.e. that the proto-fragment is optically thin to its own cooling radiation, and substitute from Eqns. (\ref{EQN:tORB}), (\ref{EQN:FBB}), (\ref{EQN:kappaR}) and (\ref{EQN:GAMMIE2}), we obtain
\begin{eqnarray}
\frac{45c_{_{\rm L}}^2h^3\Omega}{8\pi^6\bar{m}^4\kappa_{_{\rm c}}c_{_{\rm S}}^{10}}&\la&1\,.
\end{eqnarray}

Using Eqn, (\ref{EQN:OmegaR}) to replace $\Omega$, this reduces to
\begin{eqnarray}\nonumber
R\,\ga\,R_{_{\,\rm MIN}}^{^{\rm OPT.THIN}}&\!\sim\!&\left(\!\frac{3c_{_{\rm L}}\bar{m}GM_\star c_{_{\rm S}}}{5\kappa_{_{\rm c}}^2}\!\right)^{\!1/3}\frac{15h^2c_{_{\rm L}}}{4\pi^4\bar{m}^3c_{_{\rm S}}^7}\\\nonumber
&\!\sim\!&\,70\,{\rm AU}\left(\!\frac{M_\star}{{\rm M}_{_\odot}}\!\right)^{\!1/3}\left(\frac{c_{_{\rm S}}}{0.2\,{\rm km}\,{\rm s}^{-1}}\right)^{\!-20/3},\\
\end{eqnarray}
which might seem to imply that proto-fragments can condense out at any radius where the disc is optically thin, provided the gas is hot enough (sufficiently large $c_{_{\rm S}}$).

However, if the optically thin limit is to obtain, the surface-density must not be too large,
\begin{eqnarray}\nonumber
\Sigma\;\,\la\;\,\Sigma_{\tau=1}&\sim&\frac{2}{\kappa_{_{\rm c}}c_{_{\rm S}}^4}\\\label{EQN:Sigma_tau=1}
&\sim&10\,{\rm g}\,{\rm cm}^{-2}\left(\frac{c_{_{\rm S}}}{0.2\,{\rm km}\,{\rm s}^{-1}}\right)^{-4}.
\end{eqnarray}
Consequently the initial radii and masses of proto-fragments must satisfy
\begin{eqnarray}\nonumber
r_{_{\rm FRAG}}\;\ga\;r_{\tau=1}&\sim&\frac{\kappa_{_{\rm c}}c_{_{\rm S}}^6}{2\pi G}\\\label{EQN:r_tau=1}
&\sim&14\,{\rm AU}\left(\frac{c_{_{\rm S}}}{0.2\,{\rm km}\,{\rm s}^{-1}}\right)^6,\\\nonumber
m_{_{\rm FRAG}}\;\ga\;m_{\tau=1}&\sim&\frac{\kappa_{_{\rm c}}c_{_{\rm S}}^8}{2\pi G^2}\\\label{EQN:m_tau=1}
&\sim&0.001\,{\rm M}_{_\odot}\left(\frac{c_{_{\rm S}}}{0.2\,{\rm km}\,{\rm s}^{-1}}\right)^8.\hspace{1.0cm}
\end{eqnarray}

A proto-fragment with radius and mass obeying these constraints (Eqns. \ref{EQN:r_tau=1} and \ref{EQN:m_tau=1}) {\it marginally}, can only avoid being disrupted by the tidal field of the central protostar if it is formed at
\begin{eqnarray}\nonumber
R\;\ga\;R_{\tau=1}^{^{\rm TIDAL}}&\sim&\left(\frac{\kappa_{_{\rm c}}^2M_\star c_{_{\rm S}}^{10}}{(4\pi)^2 G}\right)^{1/3}\\\label{EQN:R_TIDAL}
&\sim&70\,{\rm AU}\left(\!\frac{M_\star}{{\rm M}_{_\odot}}\!\right)^{\!1/3}\left(\frac{c_{_{\rm S}}}{0.2\,{\rm km}\,{\rm s}^{-1}}\right)^{10/3}\!;\hspace{0.8cm}
\end{eqnarray}
and, if it obeys them {\it more conservatively}, it has to form even further out from the central protostar.

Evidently, it is hard for optically thin proto-fragments to condense out if the gas in the disc is much hotter than $10\,{\rm K}$ ($c_{_{\rm S}}\sim 0.2\,{\rm km}\,{\rm s}^{-1}$). For example, a modest temperature increase to $20\,{\rm K}$ ($c_{_{\rm S}}\sim 0.3\,{\rm km}\,{\rm s}^{-1}$) increases the critical tidal radius (Eqn. \ref{EQN:R_TIDAL}) to $R_{\tau =1}^{^{\rm TIDAL}}\sim 200\,{\rm AU}$, and the minimum proto-fragment radius (Eqn. \ref{EQN:r_tau=1}) to $r_{\tau =1}^{^{\rm OPT.THIN}}\sim 110\,{\rm AU}$. Even setting aside the assumption implicit in our analysis, that a proto-fragment be much smaller than the distance to the central primary protostar, this would require the disc to extend to $\sim 300\,{\rm AU}$, with surface-density of order $10\,{\rm g}\,{\rm cm}^{-2}$, and hence a rather large disc mass $M_{_{\rm DISC}}\ga {\rm M}_{_\odot}$. Any further temperature increase would make fragmentation of an optically thin part of the disc very unlikely.

In other words, at radii where a disc has sufficiently low surface-density and/or sufficiently low temperature to be optically thin, proto-fragments are so extended, and hence so susceptible to tidal disruption by the central protostar, that they are unlikely to survive -- unless they are far from the central protostar, i.e. at radii satisfying Eqn. (\ref{EQN:R_TIDAL}).  Eqn. (\ref{EQN:R_TIDAL}) defines the sweet spot for fragmentation of optically thin regions in a disc. Consideration of the extents of observed discs ($R\la 300\,{\rm AU}$) then implies that fragmentation is only likely if the temperature is low, $T\sim 10\;{\rm to}\;20\,{\rm K}$ ($c_{_{\rm S}}\sim 0.2\;{\rm to}\;0.3\,{\rm km}\,{\rm s}^{-1}$) at radii $R\ga 70\,{\rm AU}$.

\subsection{The optically thick limit}%

Conversely, if we adopt the second option in Eqn. (\ref{EQN:GAMMIE1}), i.e. that the proto-fragment is optically thick to its own cooling radiation, then substituting from Eqns. (\ref{EQN:tORB}), (\ref{EQN:FBB}), (\ref{EQN:kappaR}) and (\ref{EQN:GAMMIE2}), we obtain
\begin{eqnarray}\label{EQN:SigmaMAX.OPT.THICK}
\Sigma&\la&\Sigma_{_{\rm MAX}}^{^{\rm OPT.THICK}}\;\,\sim\;\,\left(\frac{2}{5h^3\kappa_{_{\rm c}}\Omega}\right)^{1/2}\frac{4\pi^3\bar{m}^2c_{_{\rm S}}}{3c_{_{\rm L}}}.\hspace{1.0cm}
\end{eqnarray}

In order to satisfy Eqns. (\ref{EQN:SigmaMIN}) and (\ref{EQN:SigmaMAX.OPT.THICK}) simultaneously, we require $\Sigma_{_{\rm MIN}}<\Sigma_{_{\rm MAX}}^{^{\rm OPT.THICK}}$, and, since both $\Sigma_{_{\rm MIN}}$ and $\Sigma_{_{\rm MAX}}^{^{\rm OPT.THICK}}$ are linear in $c_{_{\rm S}}$, we are left with
\begin{eqnarray}
\Omega&\la&\left(\frac{32\pi^8G^2\bar{m}^4}{45h^3c_{_{\rm L}}^2\kappa_{_{\rm c}}}\right)^{1/3}\,.
\end{eqnarray}
This upper limit on $\Omega$ translates into a lower limit on the radii at which an optically-thick disc can fragment,
\begin{eqnarray}\nonumber
R\;\,\ga\;\,R_{_{\rm MIN}}^{^{\rm OPT.THICK}}&\sim&\left(\frac{2025h^6c_{_{\rm L}}^4\kappa_{_{\rm c}}^2M_\star^3}{1024\pi^{16}G\bar{m}^8}\right)^{1/9} \\\label{EQN:R.MIN.OPT.THICK}
&\sim&70\,{\rm AU}\left(\frac{M_\star}{{\rm M}_{_\odot}}\right)^{1/3}\,,
\end{eqnarray}
a result that was first derived by \citet{MatzLevi2005}. 

If the optically thick limit is to obtain, the surface-density must be sufficiently large, $\Sigma\ga\Sigma_{\tau=1}$ (see Eqn. \ref{EQN:Sigma_tau=1}), and the initial radii and masses of proto-fragments must satisfy $r_{_{\rm FRAG}}\la r_{\tau=1}$ (see Eqn. \ref{EQN:r_tau=1}) and $m_{_{\rm FRAG}}\la m_{\tau=1}$ (see Eqn. \ref{EQN:m_tau=1}). We note that all three inequalities are reversed here, i.e. we have a lower limit on $\Sigma$ and upper limits on $r_{_{\rm FRAG}}$ and $m_{_{\rm FRAG}}$. This means that a proto-fragment with radius and mass obeying these inequalities {\it marginally} can only avoid being disrupted by the tidal field of the central protostar if it is formed at $R\ga R_{\tau=1}$ (see Eqn. \ref{EQN:R_TIDAL}). However, if a proto-fragment obeys the inequalities {\it more conservatively}, i.e. its surface-density significantly exceeds $\Sigma_{\tau=1}$, it can form closer to the central protostar and avoid being tidally disrupted. Specifically, it can form at $R_{_{\rm MIN}}^{^{\rm OPT.THICK}}$ (see Eqn. \ref{EQN:R.MIN.OPT.THICK}), provided the surface-density exceeds 
\[10\,{\rm g}\,{\rm cm}^{-2}\,\left(\frac{c_{_{\rm S}}}{0.2\,{\rm km}\,{\rm s}^{-1}}\right)^2\,.\]
Eqn. (\ref{EQN:R.MIN.OPT.THICK}) is not affected by changing $c_{_{\rm S}}$, so fragmentation of an optically thick disc is not possible closer to the central protostar than $R_{_{\rm MIN}}^{^{\rm OPT.THICK}}$ under any circumstance. Thus, for example, if the temperature is increased to $20\,{\rm K}$ ($c_{_{\rm S}}\sim 0.3\,{\rm km}\,{\rm s}^{-1}$), fragmentation is possible at $R_{_{\rm MIN}}^{^{\rm OPT.THICK}}$ provided the surface-density exceeds $\sim 20\,{\rm g}\,{\rm cm}^{-2}$. This implies a very massive disc, but is probably just credible.

In other words, if a disc has sufficiently high column-density and/or sufficiently high temperature at radii exceeding $R_{_{\rm MIN}}^{^{\rm OPT.THICK}}$ to be optically thick, proto-fragments can cool fast enough to condense out at any temperature. However, as the temperature increases, the surface-density required for a proto-fragment to avoid tidal disruption by the central protostar also increases, linearly with the temperature, and quickly reaches implausible values. Thus Eqn. (\ref{EQN:R.MIN.OPT.THICK}) defines the sweet spot for fragmentation of optically thick regions in a disc. Consideration of the surface-densities in the outskirts of observed discs ($\Sigma\la 20\,{\rm g}\,{\rm cm}^{-2}$) then implies that fragmentation is again only likely at low temperatures, $T\sim 10\;{\rm to}\;20\,{\rm K}$ ($c_{_{\rm S}}\sim 0.2\;{\rm to}\;0.3\,{\rm km}\,{\rm s}^{-1}$), since fragmentation at these low temperatures requires relatively modest surface-densities.

\subsection{Resum{\'e}}%

Irrespective of whether a disc is optically thin or thick, it seems that fragmentation is only likely to occur at radii $\ga 70\,{\rm AU}$, and even then only if the disc is cold, $T\la 20\,{\rm K}$. At smaller radii and higher temperatures, fragments are either unable to cool fast enough, or they are susceptible to tidal disruption by the central protostar. This would seem to offer a simple and attractive explanation for the Brown Dwarf Desert, i.e. the apparent paucity of brown dwarfs in close orbits round Sun-like stars \citep{MarcButl2000}. Disc fragmentation is most likely to occur in the region $R\sim 70\;{\rm to}\;100\,{\rm AU}$, and only if the temperature is below $20\,{\rm K}$ ($c_{_{\rm S}}\la 0.3\,{\rm km}\,{\rm s}^{-1}$).

\section{EPISODIC RADIATIVE FEEDBACK}\label{SEC:EPI}%

We assume that the luminosity of the central protostar is dominated by accretion, and hence its luminosity is given by
\begin{eqnarray}
L_\star&\sim&\frac{GM_\star\dot{M}_\star}{3{\rm R}_{_\odot}}\,.
\end{eqnarray}

If accretion is steady, we can define an accretion timescale,
\begin{eqnarray}
t_\star&=&\frac{M_\star}{\dot{M}_\star}\,,
\end{eqnarray}
where, for example, $t_\star =0.1\,{\rm Myr}$ corresponds to a $M_\star =1\,{\rm M}_{_\odot}$ protostar accreting at a rate of $\dot{M}_\star =10^{-5}\,{\rm M}_{_\odot}\,{\rm yr}^{-1}$. The luminosity can then be written as
\begin{eqnarray}\nonumber
L_\star&\sim&\frac{GM_\star^2}{3{\rm R}_{_\odot}t_\star}\\
&\sim&100\,{\rm L}_{_\odot}\left(\!\frac{M_\star}{{\rm M}_{_\odot}}\!\right)^2\left(\!\frac{t_\star}{0.1\,{\rm Myr}}\!\right)^{-1}\,.
\end{eqnarray}

The temperature structure in very young accretion discs can be complicated, but near the midplane -- where the density is highest, and proto-fragments are likely to originate -- the gas and dust are thermally coupled, and an approximate fit to the observed continuum emission from discs, as a function of the primary protostar's luminosity \citep[e.g.][]{OsteBeck1995} suggests that 
\begin{eqnarray}\nonumber
c_{_{\rm S}}\!(R)\!&\sim&\!0.6\,{\rm km}\,{\rm s}^{-1}\!\left(\!\frac{L_\star}{100\,{\rm L}_{_\odot}}\!\right)^{\!1/8}\!\left(\!\frac{R}{70\,{\rm AU}}\!\right)^{\!-1/4}\\\nonumber
&\sim&\!0.6\,{\rm km}\,{\rm s}^{-1}\!\left(\!\frac{M_\star}{{\rm M}_{_\odot}}\!\right)^{\!1/4}\!\left(\!\frac{t_\star}{0.1{\rm Myr}}\!\right)^{\!-1/8}\!\left(\!\frac{R}{70{\rm AU}}\!\right)^{\!-1/4}\!.\\
\end{eqnarray}
The results derived in Section \ref{SEC:BRO} indicate that such discs will have great difficulty fragmenting, because they are too hot, and indeed, in simulations with steady accretion, and hence steady radiative feedback \citep[e.g.][]{Offnetal2009, Stametal2011, Stametal2012, Lomaetal2014, Lomaetal2015a, Lomaetal2015b}, disc fragmentation is strongly suppressed.

However, if accretion is episodic, a large fraction of the final mass of a protostar is accreted during short-lived bursts, and in between there are long periods of low accretion. This means long periods of large $t_\star$ and low luminosity, during which the disc cools down, and disc fragmentation can occur quite routinely. Specifically, the low-accretion period must be longer than the time it takes for proto-fragments to condense out of the disc, which is of order the local orbital period, i.e. 
\begin{eqnarray}
t_{_{\rm ORB}}&\sim&\frac{2\pi}{\Omega}\;\sim\;10^3\,{\rm Myr}\left(\!\frac{M_\star}{{\rm M}_{_\odot}}\!\right)^{\!-1/2}\left(\!\frac{R}{70\,{\rm AU}}\!\right)^{\!3/2}\!.\hspace{0.9cm}
\end{eqnarray}
Observations suggest that accretion onto young protostars is indeed episodic \citep[e.g.][]{Kenyetal1990}, and that the low-accretion periods could be as long as $10^4\,{\rm yr}$ \citep{Schoetal2013}. Simulations of protostar formation in prestellar cores that invoke episodic accretion, using a phenomenological model based on the theory of \citet{ZhuZetal2009}, find that disc fragmentation is a regular occurrence and makes the main contribution to forming brown dwarfs and low-mass H-burning stars in the numbers observed \citep{Stametal2011, Stametal2012, Lomaetal2014, Lomaetal2015a, Lomaetal2015b}. Most of these brown dwarfs and low-mass H-burning stars are formed by disc fragmentation during low accretion episodes when the discs can cool down.

\section{HOW FAST MUST A PROTO-FRAGMENT LOSE ANGULAR MOMENTUM TO CONDENSE OUT?}\label{SEC:ORB}%

A proto-fragment can also only condense out if it is able to lose angular momentum, on a dynamical timescale; otherwise it is likely to undergo a rotational bounce, and be sheared apart or merge with another proto-fragment. Our simulations of core collapse and fragmentation (see Section \ref{SEC:SIM}) suggest that discs routinely experience impulsive perturbations. Local patches with higher than average surface-density and/or lower than average spin are created stochastically by the interaction of density waves {\it in} the disc, or the delivery {\it onto} the disc of fresh material by an anisotropic accretion stream. The tidal effect of the central protostar will then first extrude, and secondly torque, a proto-fragment, resulting in an exchange of angular momentum between the spin and the orbit of the proto-fragment. 

To evaluate the timescale on which this happens, consider a proto-fragment with radius $r$ and mass $m$, spinning at angular speed $\Omega$ (see Eqn. \ref{EQN:OmegaR}). The tidal acceleration due to the primary protostar, which acts to extrude the proto-fragment, is 
\begin{eqnarray}
\ddot{r}_{_{\rm TIDAL}}&\sim&\frac{2GM_\star r}{R^3}\,.
\end{eqnarray}
However, because the proto-fragment is spinning, this tidal acceleration only acts coherently (i.e. in approximately the same direction in a frame rotating with the proto-fragment, thereby extruding the proto-fragment) for a time $t_{_{\rm RADIAN}}\sim \Omega^{-1}$. During $t_{_{\rm RADIAN}}$, the primary protostar moves through one radian, as seen in a frame rotating with the proto-fragment. Consequently, after $t_{_{\rm RADIAN}}$ the tidal acceleration due to the primary protostar ceases to be even approximately aligned with the elongation it has caused, and the elongation saturates. Thus, the elongation is of order 
\begin{eqnarray}
\Delta r&\sim&\frac{\ddot{r}_{_{\rm TIDAL}}t_{_{\rm RADIAN}}^2}{2}\;\,\sim\;\,r\,,
\end{eqnarray}
and the fragment has an aspect ratio $\sim 2$.

The torque acting to spin down the elongated proto-fragment is
\begin{eqnarray}
\tau_{_{\rm SPIN\,DOWN}}&\sim&\frac{GM_\star m\Delta r^2}{R^3}\;\,\sim\;\,mr^2\Omega^2\,,
\end{eqnarray}
and since the moment of inertia of the proto-fragment is $I_{_{\rm FRAG}}\sim mr^2$, the time it takes to significantly reduce the spin of a proto-fragment is
\begin{eqnarray}
t_{_{\rm SPIN\,DOWN}}&\sim&\frac{I_{_{\rm FRAG}}\Omega}{\tau_{_{\rm SPIN\,DOWN}}}\;\,\sim\;\,t_{_{\rm RADIAN}}\,.
\end{eqnarray}
The net time taken to first extrude and then spin down the proto-fragment is therefore of order
\begin{eqnarray}
t_{_{\rm ANG.MOM.LOSS}}&\sim&2\,t_{_{\rm RADIAN}}\,,
\end{eqnarray}
i.e. about one third of an orbital period. If we allow that our calculation has probably overestimated somewhat the efficiency of the processes extruding and then torquing the proto-fragment, this is of order one orbital period.

This approximate calculation suggests that, in the situation where a proto-fragment initially spins at the same angular speed as it orbits the primary protostar (i.e. when a proto-fragment condenses out of an approximately Keplerian disc), the tide of the primary protostar can, in about one orbital period, induce an elongation in the proto-fragment, and then torque the elongated proto-fragment, thereby reducing its spin sufficiently to enable it to overcome rotational support and start to condense out. This timescale is essentially the period of epicyclic pulsations of the proto-fragment, and defines the maximum time available for the proto-fragment to also lose some of its thermal energy -- as assumed in Section \ref{SEC:GAM}.

\section{CONCLUSIONS}\label{SEC:CON}%

We have introduced and discussed the factors and processes that may influence the formation and gravitational fragmentation of accretion discs around newly-formed protostars forming in typical prestellar cores:
\begin{itemize}
\item We have outlined procedures for inferring, in a statistical sense, the intrinsic three-dimensional structures of prestellar cores, from dust-continuum and molecular-line observations; and how these can be used to construct initial conditions for simulations of core collapse and fragmentation.
\item We have presented the results of simulations of core collapse using initial conditions constructed in this way on the basis of detailed observations of Ophiuchus. These simulations suggest that the formation of brown dwarfs and low-mass H-burning stars requires disc fragmentation, which in turn requires (i) that a significant fraction ($\ga 0.3$) of the turbulent energy in a core be in solenoidal modes, and (ii) that accretion onto the primary protostar at the centre of the disc -- and hence also its radiative feedback -- be episodic, with a duty-cycle $\ga 3,000\,{\rm yrs}$. Typically, in the simulations using Ophiuchus-like cores, each core spawns 4 or 5 stars.
\item We have stressed that, even though the nonthermal energy (interpreted here as turbulence) in the cores in Ophiuchus is low, typically trans-sonic, the flows delivering matter into the central region where the protostars form are very irregular. This has the consequence that discs are often not well aligned with the spins of the protostars they surround, and discs are subject to perturbations due to lumpy, anisotropic inflows.
\item We have shown that, if the mapping from cores into protostars is statistically self-similar -- which it must be if the shape of the Initial Mass Function is to be inherited from the shape of the Core Mass Function -- then a typical core must spawn between 4 and 5 protostars -- in excellent agreement with the predictions of our simulations of Ophiuchus-like cores. Additionally, these protostars should have a relatively broad range of masses (interquartile mass-ratio $\sim 2$), and there should be a modest preference for the more massive protostars to end up in multiples.
\item We have presented a simple derivation of the Toomre Criterion for gravitational instability in a disc, and formulated the associated conditions on the speed with which a proto-fragment must lose thermal energy and angular momentum to condense out.
\item We have shown that the need for a proto-fragment to lose thermal energy on a dynamical timescale (the Gammie Criterion) converts into a lower limit on the radius at which a proto-fragment can condense out, which might explain the Brown Dwarf Desert.
\item We have explained why episodic feedback, with a duty-cycle $\ga 3,000\,{\rm yr}$, is required if a disc is to become sufficiently cool to fragment. We suggest that, without disc fragmentation, it is hard to form the observed numbers of brown dwarfs and low-mass hydrogen-burning stars.
\item We have shown that there is a sweet spot at which disc fragmentation, and hence the formation of brown dwarfs and low-mass hydrogen-burning stars, is most likely to occur, namely radii $70\,{\rm AU}\la R\la 100\,{\rm AU}$, and temperatures $10\,{\rm K}\la T\la 20\,{\rm K}$.
\item We have shown that tidal interactions between a proto-fragment and the primary protostar at the centre of the disc define a timescale (essentially the epicyclic period) which is the maximum time available for the proto-fragment to lose thermal energy.
\end{itemize}


\begin{acknowledgements}
APW  and OL both gratefully acknowledges the support of a consolidated grant (ST/K00926/1) from the UK's Science and Technology Funding Council (STFC).
\end{acknowledgements}


\bibliographystyle{apj}
\bibliography{antsrefs}

\end{document}